%% file: ms2_arxiv.tex
\documentclass[12pt,preprint]{aastex}

\begin{document}



\title {The Microlensing Properties of a Sample of 87 Lensed Quasars}


\author{A. M. Mosquera$^{1}$ and C. S. Kochanek$^{1,2}$}

\bigskip

\affil{$^{1}$Department of Astronomy, The Ohio State University, 140 West 18th Avenue, Columbus, 
OH 43210, USA}
\affil{$^{2}$ Center for Cosmology and Astroparticle Physics, The Ohio State University,
191 West Woodruff Avenue, Columbus, OH 43210, USA}


\begin{abstract}

Gravitational microlensing is a powerful tool for probing the physical 
properties of quasar accretion disks and properties of the lens galaxy such as its 
dark matter fraction and mean stellar mass. Unfortunately the number of lensed 
quasars ($\sim 90$) exceeds our monitoring capabilities. Thus, estimating 
their microlensing properties is important for identifying good microlensing 
candidates as well as for the expectations of future surveys. In this 
work we estimate the microlensing properties of a sample of 87 lensed quasars.
While the median Einstein radius crossing time scale is $20.6$ years, the median source 
crossing time scale is $7.3$ months. Broadly speaking, this means that on $\sim 10$ year 
timescales roughly half the lenses will be quiescent, with the source in 
a broad demagnified valley, and roughly half will be active with the 
source lying in the caustic ridges. We also found that the location of the lens 
system relative to the CMB dipole has a modest effect on  microlensing 
timescales, and in theory microlensing could be used to confirm the kinematic 
origin of the dipole. As a corollary of our study we analyzed the accretion 
rate parameters in a sub-sample of 32 lensed quasars. At fixed black hole mass, 
it is possible to sample a broad range of luminosities (i.e., Eddington factors) if 
it becomes feasible to monitor fainter lenses.

\end{abstract}


\section{Introduction}

The presence of compact objects, primarily stars and white dwarfs, close to the line of sight 
towards lensed quasar images can induce uncorrelated flux variations through gravitational 
microlensing. The relative motions of the quasar, the lens, its stars and 
the observer change the geometrical configuration of these microlenses, leading to 
changes in the total magnification. These fluctuations, suggested initially by \cite{chang79} 
and \cite{Gott}, were first detected by \cite{irwin89} in the quadruple 
lensed quasar Q2237+0305 \citep{Huchra85}. Since then gravitational microlensing has been detected 
in many other systems \citep[see review by][]{joachim-saas06}, and in the last 5 years the field 
has grown rapidly.

Microlensing depends on the size and structure of the source, the relative velocities of the 
components, the mass of the microlenses and the surface density and shear of the lens galaxy 
near the images. In particular, a range of approaches 
have been used to examine the geometry and properties of quasar accretion disks,
which are very likely to be microlensed because their sizes are similar or smaller than microlensing 
length scales. For example the amplitude of microlensing fluctuations are governed by the size of the source, 
so microlensing light curves can be analyzed to infer the spatial structure of the disk 
\citep[e.g.,][]{Wyithe2000, Wyithe2002, CK04}. This can be done by directly modeling the lightcurves, 
and the most advanced models include the motion of individual stars which allows explorations 
of the shape and orientation of the disk \citep{PK1,PK2}. In addition, if intrinsic quasar variations 
are due to changes in the area of the accretion disk, there is the possibility of measuring its effects 
on the microlensing signal \citep{jeff2010, dexter2011}. 
The continuum sizes estimated with microlensing techniques were used by \cite{morgan10} to determine 
the relationship between the accretion disk size and the black hole (BH) mass, and the scaling 
they found is consistent with the predictions of the thin disk theory ($R \propto M^{2/3}_\mathrm{{BH}}$). 
The microlensing sizes, however, seem to be larger than estimates based on either the 
observed fluxes or black hole mass measurements \citep{pooley07,morgan10}.

Measuring the amplitude of microlensing as a function of wavelength constrains the 
temperature profile of the optical emitting region of the disk. Since quasar accretion disk models 
\citep[e.g.,][]{blaes04} predict cooler temperatures for more distant, larger parts 
of the disk, different microlensing magnifications are expected for different temperatures 
(i.e. at different wavelengths). This wavelength dependence, also known as chromatic microlensing, 
has been observed and analyzed in many systems \citep{Lutz93, Lutz95,Olaf03,timo2008,bate08,eigenbrod08, 
poindexter08, Floyd2009, ana09, ana10,jeff20102, evencio10} and, except for the results 
found by \cite{Floyd2009}, the temperature profiles are generally consistent with simple thin disk 
models \citep{ss1973} or with shallower temperature profiles that would help to explain the 
size discrepancies \citep[see][]{morgan10}. 
 
Microlensing has also been used to study the geometry of the non-thermal emission regions. For 
example, the production of the observed X-ray radiation from the disk is currently described by 
two general scenarios \citep[e.g.,][]{RC2000,aneta1,aneta2} but it is still not well understood. Several 
microlensing studies combining optical and X-ray flux ratios 
\citep{pooley07, morgan08, chartas09, dai_csk, jeff20102} suggest that the X-ray emission 
originates in a compact region close to the central BH. However, further studies are needed to 
verify these results and to determine the scaling of the X-ray emission 
region with energy and BH mass. The spatial structure of the broad line region (BLR) of quasars 
has also been explored using microlensing \citep[e.g.,][]{Lewis98,Cristina02,Sluse07,Sluse11,ODowd10}.

Finally, microlensing has also been used to study the distribution of matter in the lensing galaxies. 
\cite{morgan08}, \cite{dai10}, and \cite{bate11} considered individual systems, while \cite{evencio09} and 
\cite{pooley09} surveyed multiple lens galaxies. These studies consistently favor models where 
the surface mass density is dominated by dark matter and stars represent only $\sim 10\%$ of the local matter. 
The exception is the lens Q~2237+030, where the images are seen through the bulge of a 
nearby spiral galaxy and the models correctly find that the surface density is dominated 
by stars.

The number of known lensed quasars\footnote{Lens quasars have been discovered in many cases serendipitously, 
and in other cases they were found in a broad different range 
of lens/quasar surveys. A compact summary of these surveys and the different methodologies applied to find the lenses can be found 
in \cite{CKsaas06}.} ($\sim 90$) is presently much larger than our capacity to regularly monitor for microlensing variability. 
While some estimates can be made from ``static'' measurements as a function of wavelength \cite[e.g.,][]{Floyd2009, ana10,evencio10}, 
these require strong priors that are subject to systematic problems and cannot probe all the physics involved. 
Given this problem, it seems useful to survey the microlensing properties of the known lenses to identify 
those that may have shorter timescales, larger amplitudes, or other properties that make them better 
(or worse) targets. With that aim, we have joined together the lensed quasars from the CfA-Arizona Space 
Telescope LEns Survey \citep [CASTLES;][]{castles} of gravitational lenses\footnote {http://www.cfa.harvard.edu/castles/} 
and the Sloan Digital Sky Survey Quasar Lens Search \citep [SQLS;][]{oguri06}\footnote{http://www-utap.phys.s.u-tokyo.ac.jp/$\sim$sdss/sqls/}, 
and we have estimated the microlensing properties of 87 systems. In Section \ref{sec2} we review the theoretical background 
on microlensing properties, and the selection criteria for the lensed quasars used in our analysis. 
In Section \ref{sec3} we interpret the results and discuss their the main implications. Throughout this work we assume 
$\Omega_m=0.3$, $\Omega_{\Lambda}=0.7$, and  $H_0=72$ km sec$^{-1}$ Mpc$^{-1}$.

\section{Procedure}\label{sec2}

The magnification pattern caused by the stars and 
compact objects in the lens galaxy  has a characteristic 
scale set by the Einstein radius of the microlenses, $R_E$. 
Microlensing fluctuations will be significant if the 
characteristic source size, $R_S$, satisfies $R_S\lesssim R_E$, 
and their amplitude will be controlled by $R_S/R_E$, with smaller 
ratios leading to larger amplitudes. The timescales for microlensing 
variations to occur will be given by the characteristic times $t_E=R_E/v$ 
and $t_S=R_S/v$, where $v$ is the effective transverse velocity of the source. 

\subsection {Microlensing length scales}

The microlensing regime is governed by the Einstein radius of the microlenses, which is defined on 
the source plane by
\begin{equation}\label{re}
R_E= D_{OS}\left[ \frac{4G \langle M \rangle}{c^2} \frac{D_{LS}}{D_{OL} D_{OS}}\right]^{1/2},
\end{equation}
where $G$ is the gravitational constant, $c$ is the speed of light, $\langle M \rangle$ is the mean mass 
of the compact objects, and  $D_{OL}$, $D_{OS}$, $D_{LS}$ are the angular diameter distances between the 
observer, the lens, and the source.

The other length scale on which microlensing depends is the size of the source, 
$R_S$, since fluctuations are expected to be important when $R_S\lesssim R_E$. In the case 
of lensed quasars, the accretion disk should generally  be sufficiently compact to satisfy 
this criterion for typical microlens masses and standard source and lens redshifts. The accretion disk size $R_S$ 
can be estimated assuming a simple thin-disk model \citep{ss1973}, from either the measured flux 
(Equation \ref{rflux}), or the central BH mass, $M_{\mathrm{BH}}$ (Equation \ref{rmbh}). For a 
thin disk emitting as a black body with temperature profile $T\propto R^{-3/4}$, and (safely) ignoring relativity 
and inner edge effects, the radius where the disk temperature matches the rest wavelength 
of the observations ($kT=h_pc/\lambda_{\mathrm{rest}}$) is given by
\begin{equation}\label{rflux}
R^{\mathrm{flux}}_{\lambda} \simeq \frac{3.4 \times 10^{15}}{\sqrt{\cos i}} \frac {D_{OS}}{r_H} 
\left(\frac{ \lambda}{\mu m} \right)^{3/2} \left(\frac{\mathrm{zpt}}{3631 \ \mathrm{Jy}}\right)^{1/2}
10^{-0.2 \ (m-19)} \ h^{-1}\ \mathrm {cm},
\end{equation}
where $D_{OS}/r_H$ is the angular diameter distance to the quasar in units of the Hubble radius, 
$i$ is the disk inclination angle, and $m$ is  magnification-corrected magnitude. 
We have normalized Equation \ref{rflux} to the zero point (zpt) of AB magnitude system \citep{ABmag}, but have primarily applied it to 
the observed I band magnitudes ($\lambda=0.814 \ \mu$m  and zpt$=2409$ Jy). We refer to this as the ``flux size'' 
$R_I$ of the quasar. For the same disk model, the disk size can be estimated from the mass of the 
black hole $M_{\mathrm{BH}}$ and the luminosity $L$ as
\begin{equation}\label{rmbh}
R^{\mathrm{theory}}_{\lambda}= 9.7\times 10^{15}  \left(\frac{\lambda_{\mathrm{rest}}}{\mu m} \right)^{4/3}\left(\frac{M_{\mathrm{BH}}} 
{10^9 M_{\odot}}\right)^{2/3}\left( \frac{L}{\eta L_E}\right)^{1/3} \ \mathrm {cm},
\end{equation}
where $L/L_E$ is the luminosity in units of the Eddington luminosity, and $\eta$ is the accretion 
efficiency. As discussed by \cite{morgan10}, these two estimates should in principle yield the 
same disk size. However, for typical accretion rate values, $L/L_E \sim 1/3$ and $\eta=0.1$ 
\citep[e.g.,][]{juna06, hopkins09, Lutz10}, the flux sizes are significantly smaller than the 
theoretical sizes \citep[e.g.,][]{collin02}. This can be rectified by assuming higher efficiencies, 
or very low luminosities, but this leads to problems for models of black hole growth 
\citep{weinberg03,shankar04, shankar09}. \cite{dexter2011} argue that temperature fluctuations in the disk can 
increase the effective disk size compared to Equation \ref{rflux} and rectify these differences. While 
uncertainties in the BH mass and magnification measurements also affect these estimates, they are generally 
less important and too small to solve the problem.

Since black hole masses have been measured for only 32 of the known lensed quasars (see below), in most 
of this work we will refer to the source disk size, $R_S$, as the one derived from the I-band flux 
in Equation \ref{rflux} (unless otherwise stated). We adopted a mean inclination of $\langle \cos i \rangle=1/2$ in all cases, and the 
magnification-corrected magnitudes were calculated from the observed magnitudes of the CASTLES and the SQLS surveys using 
either already existing lens models (see references in Table \ref{table1}) or corrected by magnifications estimated from simple singular 
isothermal sphere (SIS) plus shear models. In the case of the systems B0850$+$054 and Q1208+101, where the position of the lens galaxy 
is unknown, we adopted a mean magnification $\langle\mu\rangle=4$ \citep{turner84}. For those systems lacking I-band measurements, but 
observed at other wavelengths, the size in Equation \ref{rflux} was calculated for that band and then converted to $R_I$ assuming the 
size-wavelength 
scaling of the thin disk model $R_{\lambda}\propto \lambda^{4/3}$. We focused on a fixed observed rather than rest wavelength because 
it is more closely related to observations. \cite{pooley07} and \cite{morgan10} found that these flux sizes generally underestimate 
the size inferred from microlensing by a factor of $\sim$ 2~-~3.

We have also estimated the sizes of the BLR, $R_{\mathrm{BLR}}$, although they are less likely 
to be microlensed due to their bigger extent \cite [e.g.,][]{Bentz2006}. We used the H$\beta$ BLR 
size-luminosity relationship, 
\begin{equation}\label{eq_BLR}
\mathrm{log}_{10}(R_{\mathrm{BLR}})=K + \alpha \ \mathrm{log}_{10}(\lambda L_{\lambda}(5100 \mathrm {\AA})),
\end{equation}
with $\alpha=0.519$ and $K=-21.3$ \citep{Bentz2009}. The mean luminosity at rest-frame 5100~\AA, 
$\lambda L_{\lambda}(5100 \mathrm {\AA})$, was calculated using a quasar spectral energy distribution 
template \citep{Roberto10b} to scale the I-band flux to an estimate of the rest-frame flux at 5100 \AA.

\subsection {Microlensing timescales}

The characteristic timescales for microlensing variability combine the length scales with the 
expected effective velocity $v$ of the source. If $R_S\lesssim R_E$, microlensing fluctuations 
will certainly occur on the timescale $t_E=R_E/v$, at which the source moves an Einstein radius. If the 
source is compact and moving through active regions of the caustic networks, magnification changes are 
expected on a shorter timescale, $t_S=R_S/v$, 
corresponding to the time it takes the source to cross its own size, $R_S$. The effective velocity has 
three components \citep{Kayser86}, corresponding to the motions of the observer, the lens 
and the source.
We estimated an effective source plane velocity following \cite{CK04} as
\begin{equation}\label{vel_ef}
v^2_e=\left( \frac{\sigma_{\mathrm{pec}}(z_l)}{1+z_l} \frac{D_{OS}}{D_{OL}}\right)^2+\left(\frac{\sigma_{\mathrm{pec}}(z_s)}{1+z_s}\right)^2+ 
\left( \frac{v_{\mathrm{CMB}}}{1+z_l} \frac{D_{LS}}{D_{OL}}\right)^2 +
2\left( \frac{\sigma_{\ast}}{1+z_l} \frac{D_{OS}}{D_{OL}}\right)^2
\end{equation}
where $\sigma_{\mathrm{pec}}$ is the one dimensional rms galaxy peculiar velocity at redshift $z$, $v_{\mathrm{CMB}}$ is the 
projection of the  CMB dipole velocity \citep{cmb09} onto the lens plane, and $\sigma_{\ast}$ is the
velocity dispersion of the stars in the lens galaxy. We evaluated $\sigma_{\mathrm{pec}}$ 
based on the peculiar velocity models of \cite{tinker}\footnote{Their results were obtained using a subhalo abundance matching 
technique, see \cite{Conroy2006} and \cite{Wetzel2010} for details on this method.}, interpolating between their 
fiducial redshifts as log$_{10}$(1+z), with
\begin{equation}\label{spec}
\mathrm{log}_{10}\left(\frac{\sigma_{\mathrm{pec}}}{\mathrm{km\ s}^{-1}}\right)= a~\mathrm{log}_{10}(1+z)+b,
\end{equation}
where the coefficients $a$ and $b$ are given in Table \ref{coef}. 
We estimated the stellar velocity dispersion 
$\sigma_{\ast}$ from the image separation as
\begin{equation}\label{dtheta}
\Delta \theta=8\pi \left(\frac{\sigma_{\ast}}{c}\right)^2\frac{D_{LS}}{D_{OS}},
\end{equation}
since the velocities inferred from the SIS model agree well with the measured dispersions of 
lens galaxies \citep[e.g.,][]{treu09}.

 \subsection {Redshifts}

In order to calculate the microlensing timescales we need both the source and lens redshifts. 
Unfortunately  in many cases they are not known. We restricted our study to 87 lensed quasars with 
known source redshifts. Of these, 33 lack spectroscopic lens redshifts. Table \ref{table1} lists the 
systems and their properties. Where lens redshifts are missing, we used estimates based on the fundamental 
plane \citep [e.g.,][]{CK00, Rusin03}, the \cite{FJ76} relationship \citep [e.g.,][]{Keeton98, Rusin03}, 
galaxy colors \citep[e.g.,][]{kayo10}, and clear detections of strong, lower redshift absorption lines 
in the quasar spectra \citep[e.g.,][]{Lacy02}. As a final resort, we use the maximum likelihood lens 
redshift for producing a lens of separation $\Delta \theta$ from a source at redshift $z_s$ 
following \cite{ofek03}. In Table \ref{table1} we note which method was applied to estimate the unknown 
lens redshifts.

\section{Results and Discussion}\label{sec3}

\begin{figure}
\begin{center}
\vspace{0.5 cm}
\includegraphics[scale=0.8]{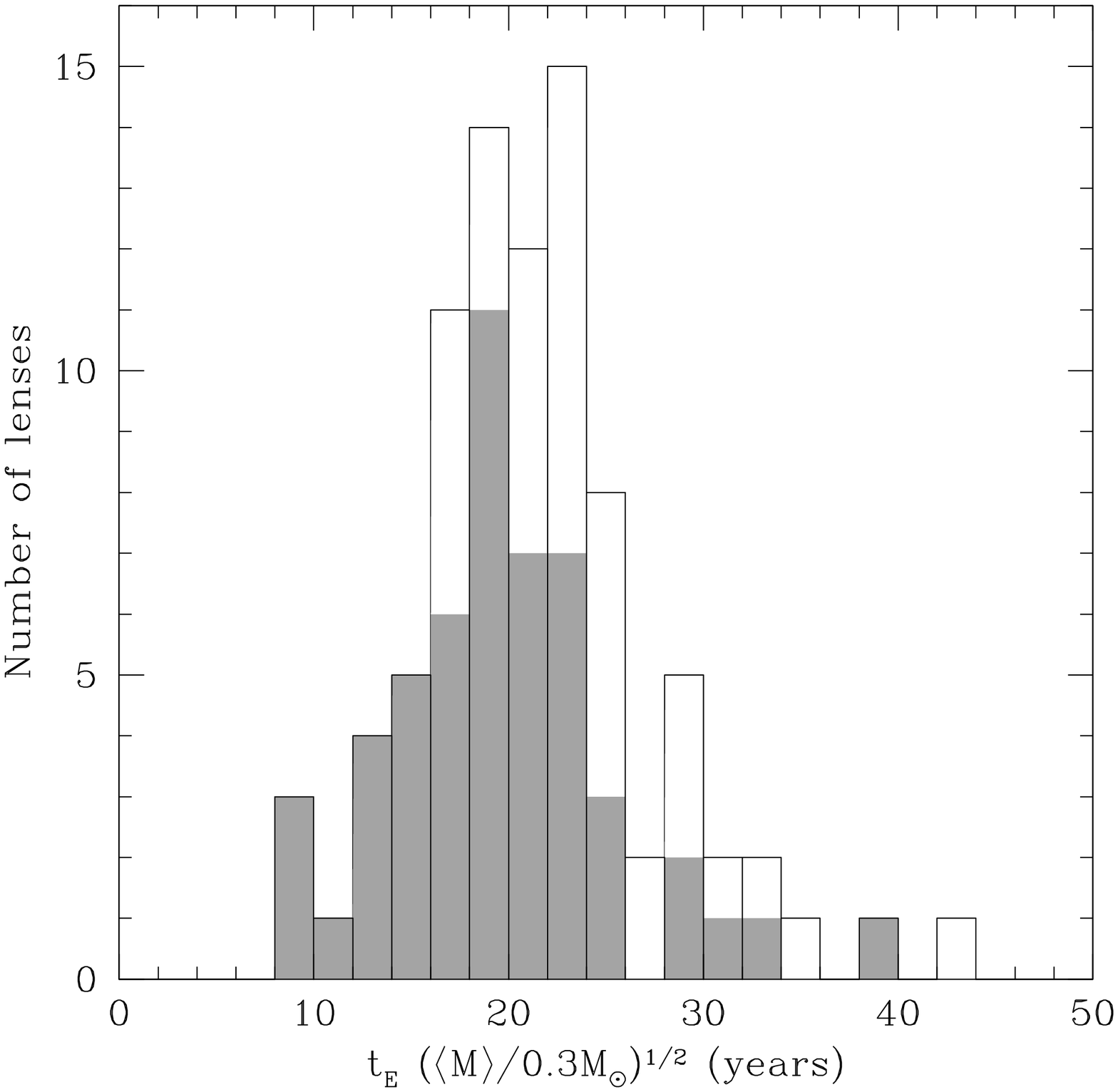}
\caption{\label{te} Histogram of the lenses as a function of the Einstein crossing timescale $t_{E}=R_E/v$. 
The gray bars exclude those systems lacking spectroscopic redshifts for the lens galaxy.}
\end{center}
\end{figure}

\begin{figure}[t]
\begin{center}
\vspace{0.5 cm}
\includegraphics[scale=0.8]{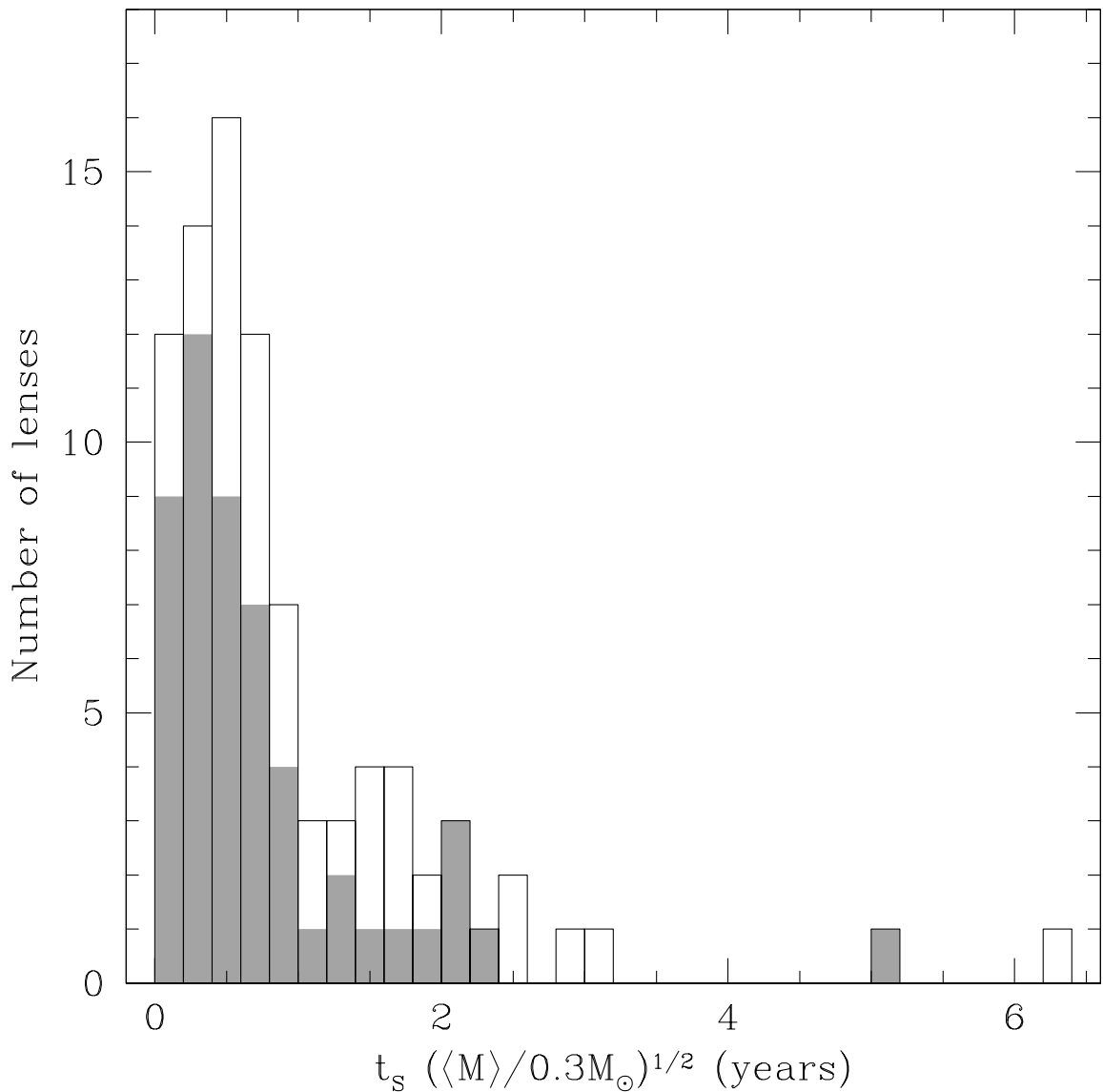}
\caption{\label{ts} Histogram of the lenses as a function of the source crossing timescale $t_{S}=R_S/v$, 
where $R_{S}$ is the size estimated from the I-band (or other optical) flux following Equation \ref{rflux}. 
The gray bars exclude those systems lacking spectroscopic redshifts for the lens galaxy. Measurements of 
the sizes with microlensing tend to be a factor of 2$-$3 larger.}
\end{center}
\end{figure}

We summarize the microlensing scale estimates assuming a mean stellar mass 
of $\langle M \rangle=0.3M_{\odot}$ in Table \ref{table1}. Figures \ref{te} and  \ref{ts} show the histograms for 
$t_E$ and $t_S$, respectively, where the gray histogram includes only the 
systems with a spectroscopic redshift for the lens galaxy. The Einstein crossing time has a peak at $\sim23$ years, 
and ranges from $\sim 8$ years (Q~2237+0305) to $\sim 44$ years (Q~1208+101). The shape of 
the histogram can be explained by combining Equations \ref{re} and \ref{dtheta} to find that
\begin{equation}
t_{E}= \left(\frac{R_E}{v}\right) \sim \ \left( \frac{\langle m\rangle}{\Delta \theta}\right)^{1/2}
\left( \frac{D_{LS}}{D_{OL}}\right)
 \left(\frac{\sigma_{\ast}}{\sigma_{\ast}+v_{\mathrm{bulk}}}\right)^{1/2}.
\end{equation}
Short timescales are rare because there is little volume at low redshifts (the $D_{LS}/D_{OL}$ term), 
and long time scales are rare because large separation lenses are rare (the $1/\Delta \theta$ term). The 
bulk velocity $v_{\mathrm{bulk}}$ matters little unless it is very large compared to the lens dispersion. The 5 
shortest Einstein crossing timescales ($8 \lesssim t_E \lesssim12$ years) are for Q0957$+$561, 
SDSSJ1004$+$4112, SDSS1029$+$2623, RXJ1131$-$1231, and Q2237$+$030, where the first three systems 
are lensed by galaxy groups or clusters having higher velocity dispersions. In the case of Q2237$+$030 
and to a lesser extent, RXJ1131$-$1231, the small lens distance $D_{OL}$ ``magnifies'' all the velocity terms in 
Equation \ref{vel_ef} to give a short time scale. 

Although the characteristic times to cross the 
Einstein radius are sometimes discouragingly large, the source crossing time scales on which we can see 
microlensing variations are considerably  shorter (Figure \ref{ts}), ranging from $\sim 1$ week 
(B2045$+$265) to $\sim 6$ years (Q1208$+$101), although in this latter case we assumed the mean magnification 
correction of a SIS model. The 5 shortest timescale systems are  PMNJ0134$-$0931, B0712$+$472, 
SDSS1011$+$0143, B1359$+$154, and B2045$+$265 ($ 1\lesssim t_S \lesssim 8$ weeks). Unfortunately, small 
sources also correspond to low luminosity sources, so these tend to be some of the hardest systems 
to monitor. If a lens is in an active region of caustics it should be microlensed on very short time 
scales and we found that $17\%$ of the lens systems have $t_S\lesssim 3$ months, $36\%$  have 
$t_S\lesssim 6$ months, and $70\%$ have $t_S \lesssim 1$ year. Even though our estimates are based on rough 
approximations, microlensing fluctuations observed in regularly monitored systems like 
HE~0435$-$1223 \citep[e.g.,][]{jeff2010,courbin10}, RXJ~1131$-$1231 \citep[e.g.,][]{Szy2010} 
and Q~2237$+$0305 \citep[e.g.,][]{ogle2237} agree well with our predictions. For those systems we 
estimated $t_S \sim 9$ months for the first and $t_S\sim 3$ months for the other two.

\begin{figure}[t]
\begin{center}
\vspace{0.5 cm}
\includegraphics[scale=0.8]{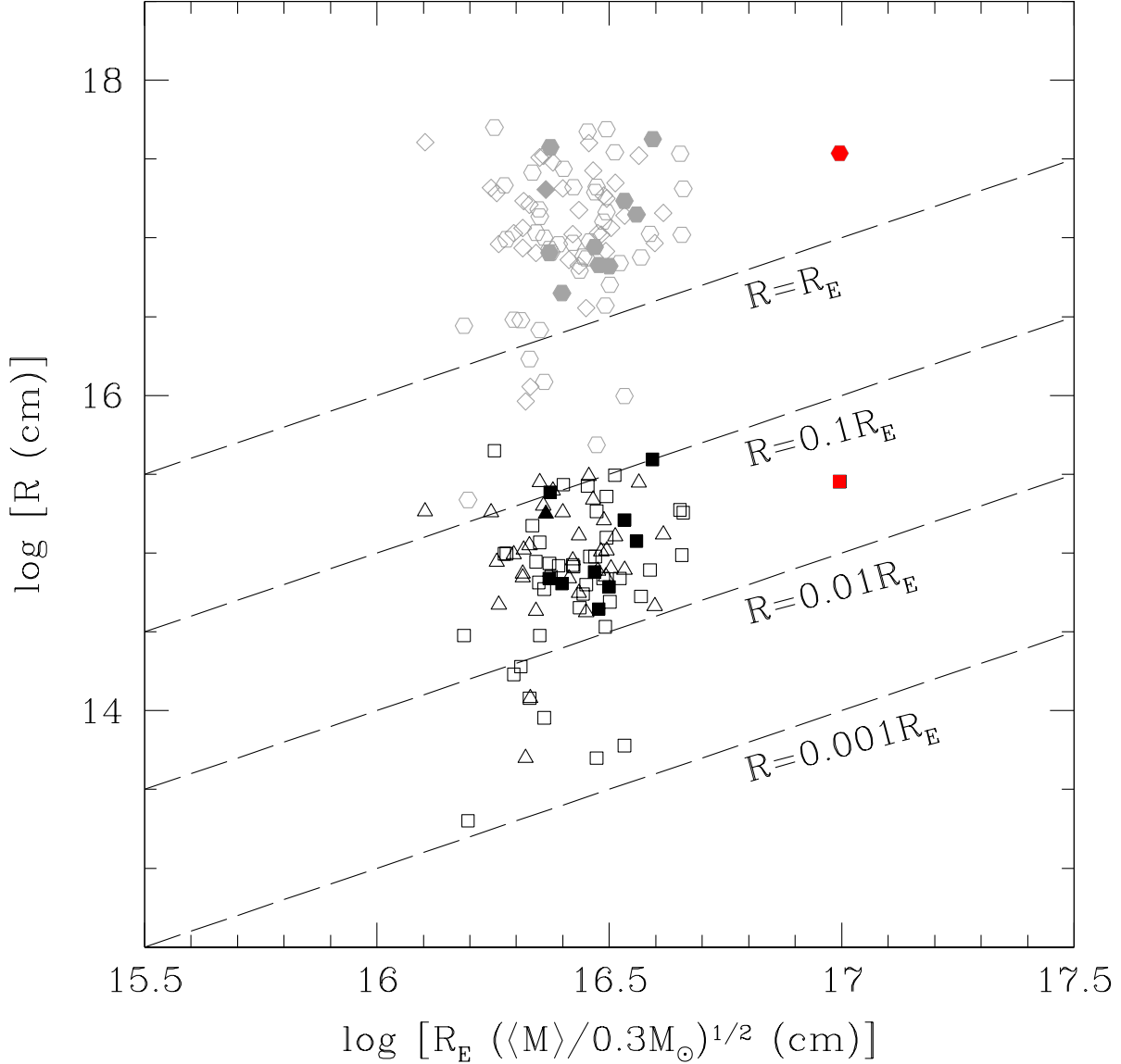}
\caption{\label{RE_RI} Distribution of lenses in source size $R$ and Einstein radius, $R_E$. 
The dashed lines mark various size ratios and microlensing fluctuations will be stronger for 
smaller values of $R/R_E$. Squares and triangles (black points) correspond to flux size measurements $R_S$
(at I-band), where squares represent systems with spectroscopic redshifts for $z_l$, and triangles those with 
only estimates. Microlensing measurements suggest $R_S$ underestimates the effective 
source sizes by factors of 2$-$3 \citep{pooley07,morgan10}. Hexagons and diamonds (gray points) 
correspond to BLR-size estimates, $R_{\mathrm{BLR}}$, for systems with and without 
spectroscopic lens redshifts, respectively. The red point corresponds to the lensed quasar Q~2237+0305.}
\end{center}
\end{figure}

\begin{figure}[t]
\begin{center}
\vspace{0.5 cm}
\includegraphics[scale=0.8]{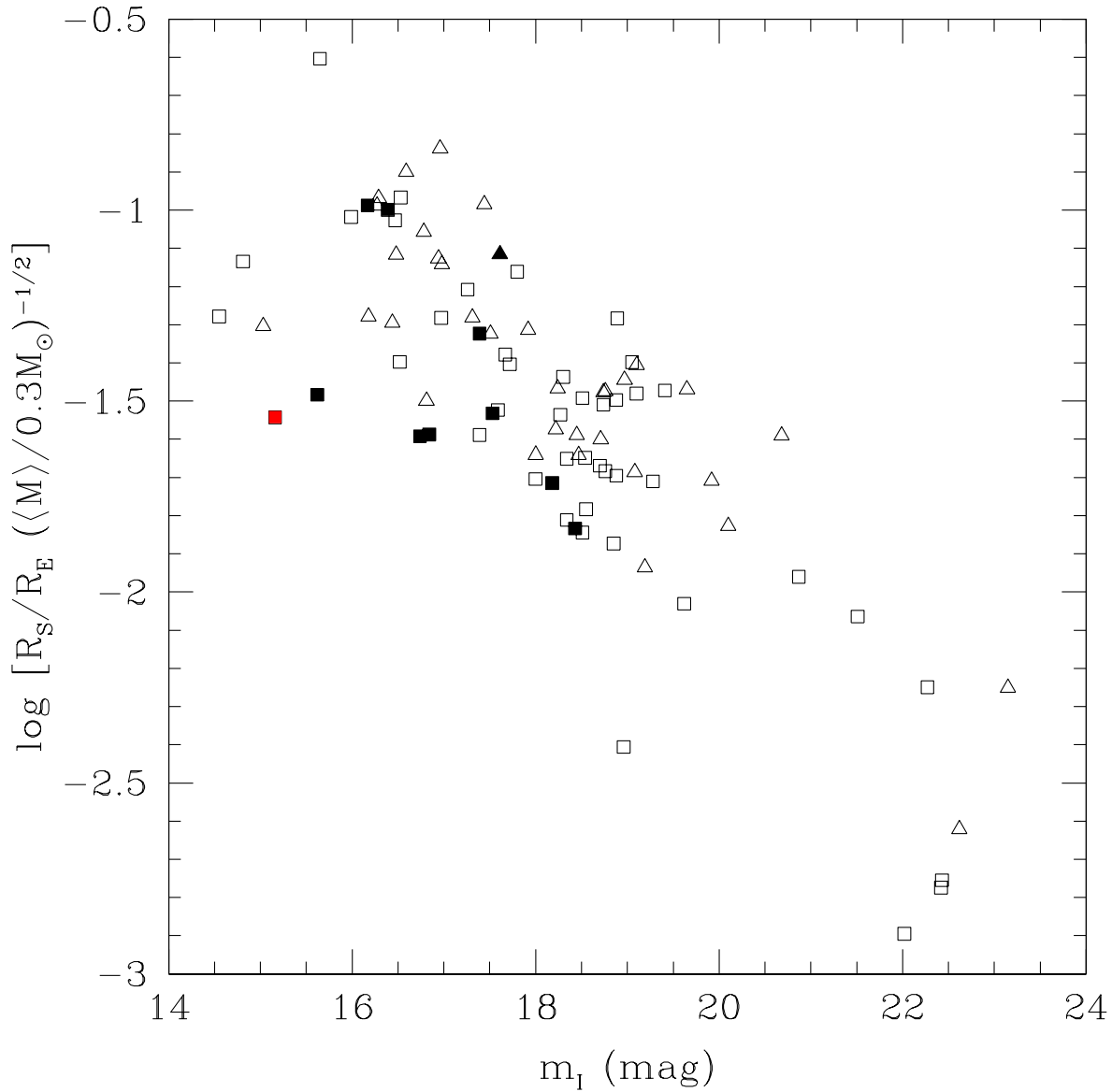}
\caption{\label{mag_coc} The ratio between the flux size and the Einstein radius, log [$R_S/R_E$], as a function of 
the observed (total) I-band magnitude, $m_I$. Squares correspond to systems with spectroscopic redshifts 
for $z_l$, while triangles indicate those with only estimates. The filled symbols indicate the 
lenses from \citet{morgan10} with microlensing size measurements, and the red square corresponds 
to Q~2237+0305. Under the assumption of Equation \ref{rflux} that the luminosity determines the disk size, the 
systems with the greatest expected level of microlensing variability will also tend to be faint.}
\end{center}
\end{figure}

The amplitudes of the microlensing fluctuations are controlled by $R_S/R_E$, with smaller ratios leading to 
higher amplitudes. For $\langle M \rangle=0.3M_{\odot}$ all the systems have $R_S \lesssim R_E$ 
(Figure \ref{RE_RI}, black points), and the ratio ranges from $\sim0.001$ (B2045$+$265) to $0.25$ 
(SBS0909$+$523). While the exact ratio depends on the assumed mean mass $\langle M \rangle$, it is unlikely 
that  $\langle M \rangle$ changes greatly between lenses, so the relative ordering of the systems is principally 
uncertain due to the uncertainties in $R_{S}$, where the largest problem is the evidence from microlensing 
of an offset between $R_S$ from Equation \ref{rflux} and microlensing measurements \citep{pooley07,morgan10}. 
In Figure \ref{RE_RI} we have marked the systems with microlensing size measurements \citep[filled squares;][]
{morgan10}, and we see that they tend to be systems with ratios $0.01\lesssim R_S/R_E \lesssim 0.1$. The 5 systems 
with the largest potential amplitudes are PMNJ0134$-$0931, B0712$+$472, SDSS1011$+$0143, B1359$+$154, and B2045$+$265. 
These are due to small sources rather than  large $R_{E}$. Figure \ref{mag_coc} illustrates the strong 
correlation of $R_{S}/R_{E}$ with the total flux of the lens that is expected given a strong correlation of the disk 
size with luminosity (Equation \ref{rflux}). If we are restricted to systems brighter than 19 mag in the $I$-band, then 
the best five systems to search for higher amplitude, shorter timescale microlensing variability are PMNJ0134$-$0931, SDSSJ0819$+$5356, 
SDSSJ1029$+$2623, SDSSJ1251$+$2935, and Q2237$+$0305. While relatively large amplitudes have been observed in Q2237$+$0305 microlensing 
fluctuations have not been observed in PMNJ0134$-$0931 (which may be problematic for other reasons, see \cite{KW2003}), 
and SDSSJ1029$+$2623 is a cluster lens where there may be few stars near the lensed images. In the case of SDSSJ0819$+$5356 
and SDSSJ1251$+$2935 modest flux ratio anomalies have been observed  and better observations are required to understand their origin.
In Figure \ref{RE_RI} we have also plotted the BLR sizes for comparison (gray points), and although most of the systems 
have $R_E\lesssim R_{\mathrm{BLR}}$ some of them would be likely to show microlensing fluctuations in the BLR. These small BLR-size 
systems have an obvious correspondence with those having smaller accretion disks.

\begin{figure}[t]
\begin{center}
\vspace{0.5 cm}
\includegraphics[scale=0.8]{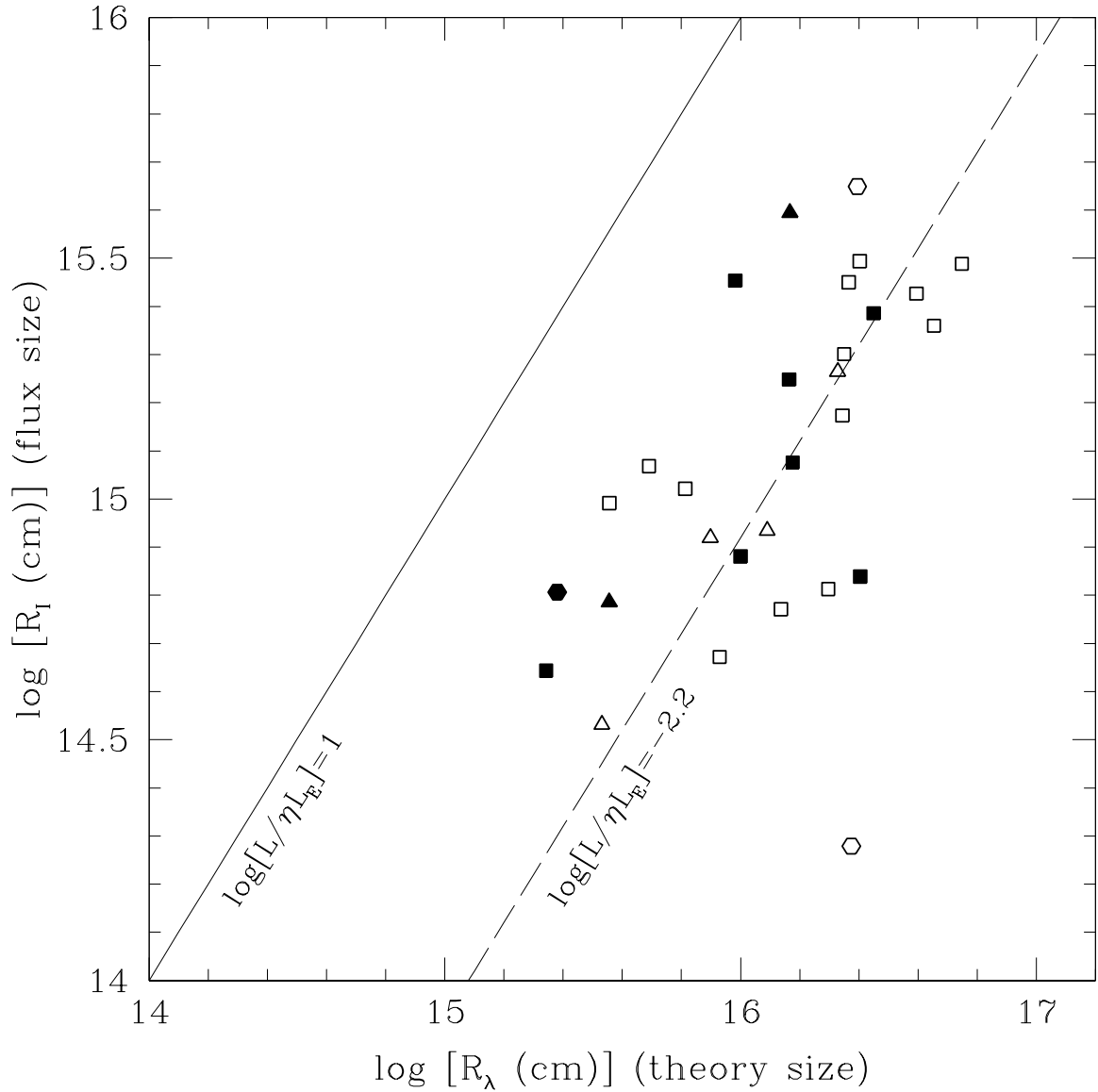}
\caption{\label{RE_RBH} Thin disk flux sizes, $R_I$ (Equation \ref{rflux}), as a function of the size 
estimates from BH masses, $R_{\lambda}$ (Equation \ref{rmbh}), assuming $\eta=0.1$ and $L/L_E=1$. 
The different symbols correspond to BH mass estimates based on  CIV (squares), H$\beta$ 
(hexagons), and MgII (triangles) line widths. The filled symbols indicate those systems with microlensing sizes. 
The solid line corresponds to the relationship log $(L/\eta L_E)=1$ and the dashed line 
corresponds to the offset of log $(L/\eta L_E)=-2.2$ needed to make the two estimates consistent.}
\end{center}
\end{figure}

\begin{figure}[t]
\begin{center}
\vspace{0.8 cm}
\includegraphics[scale=0.8]{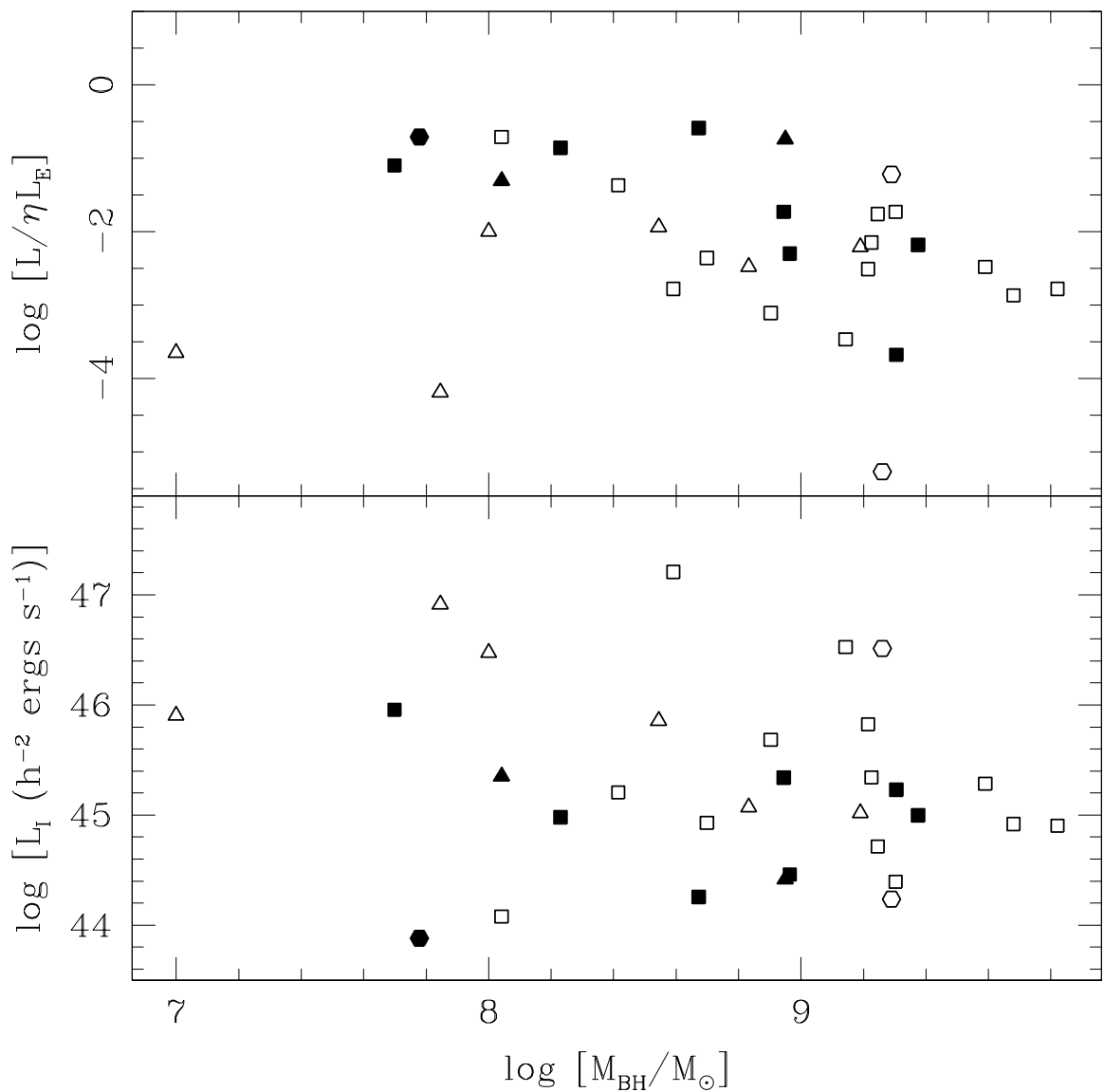}
\caption{\label{Efact_BH} Top panel: Eddington factors, log $(L/\eta L_E)$, versus BH mass, based on 
combining Equations \ref{rflux} and \ref{rmbh}. Bottom panel: Magnification-corrected I-band luminosities 
as a function of the BH mass. The different symbols 
correspond to BH mass estimates based on CIV (squares), H$\beta$ (hexagons), and MgII (triangles) 
line widths. The filled symbols correspond to those systems with microlensing sizes.}
\end{center}
\end{figure}

Since \cite{morgan10} found correlations of disk size with BH masses, the next frontier is to search for 
correlations with accretion states, particularly the Eddington factor. For those lens systems where the 
mass of the BH has been estimated from line width measurements \citep{peng06, Greene10, Roberto10}, 
we have calculated $R_{\lambda}$ for $L=L_E$ and $\eta=0.1$ based on Equation \ref{rmbh}. In Figure 
\ref{RE_RBH} we compare these to the disk size estimates from the I-band flux, $R_I$, using Equation 
\ref{rflux}. As noted by \cite{pooley07} and \cite{morgan10}, we see that most of the systems lie below the one-to-one 
relationship. The mean offset corresponds in Equation \ref{rflux} to an Eddington efficiency factor of 
log$(L/\eta L_E)\sim-2.2$. If we can use the flux size from Equation \ref{rflux} as an estimate of the disk size in 
Equation \ref{rmbh}, we can estimate the Eddington term as $(L/\eta L_E)\propto R^3_{\lambda} M^{-2}_{\mathrm{BH}}$, 
as shown in Figure \ref{Efact_BH} (top panel). The strong trend with BH mass and the presence of extreme values suggests 
that this may not be a reliable indicator of differences in accretion state at fixed BH mass. Alternatively 
we could simply examine the distribution of magnification corrected luminosity\footnote{The observed luminosity of the source in I-band 
is given by $L_I=3.88\times10^{37} h^{-2} D_{OS}^2 (1+z_s)^4 10^{0.4(m_I-19)}$ ergs s$^{-1}$, where $m_I$ is the unmagnified 
observed magnitude, and $D_{OS}$ is the angular diameter distance.} with black hole mass, as also shown 
in Figure \ref{Efact_BH} (bottom panel). It certainly seems  possible to sample a relatively broad range 
of source sizes at a fixed $M_{\mathrm{BH}}$ given the distribution of either the Eddington term estimate or the intrinsic luminosities. 
The problem again is that most of the low luminosity quasars with low Eddington factors and small $R_S$ are also faint. 
Quantitatively exploring these systems will likely require deeper surveys such as 
the Large Synoptic Survey Telescope\footnote{http://www.lsst.org/lsst/} \citep{LSST06}.

\begin{figure}[t]
\begin{center}
\vspace{0.5 cm}
\includegraphics[scale=0.8]{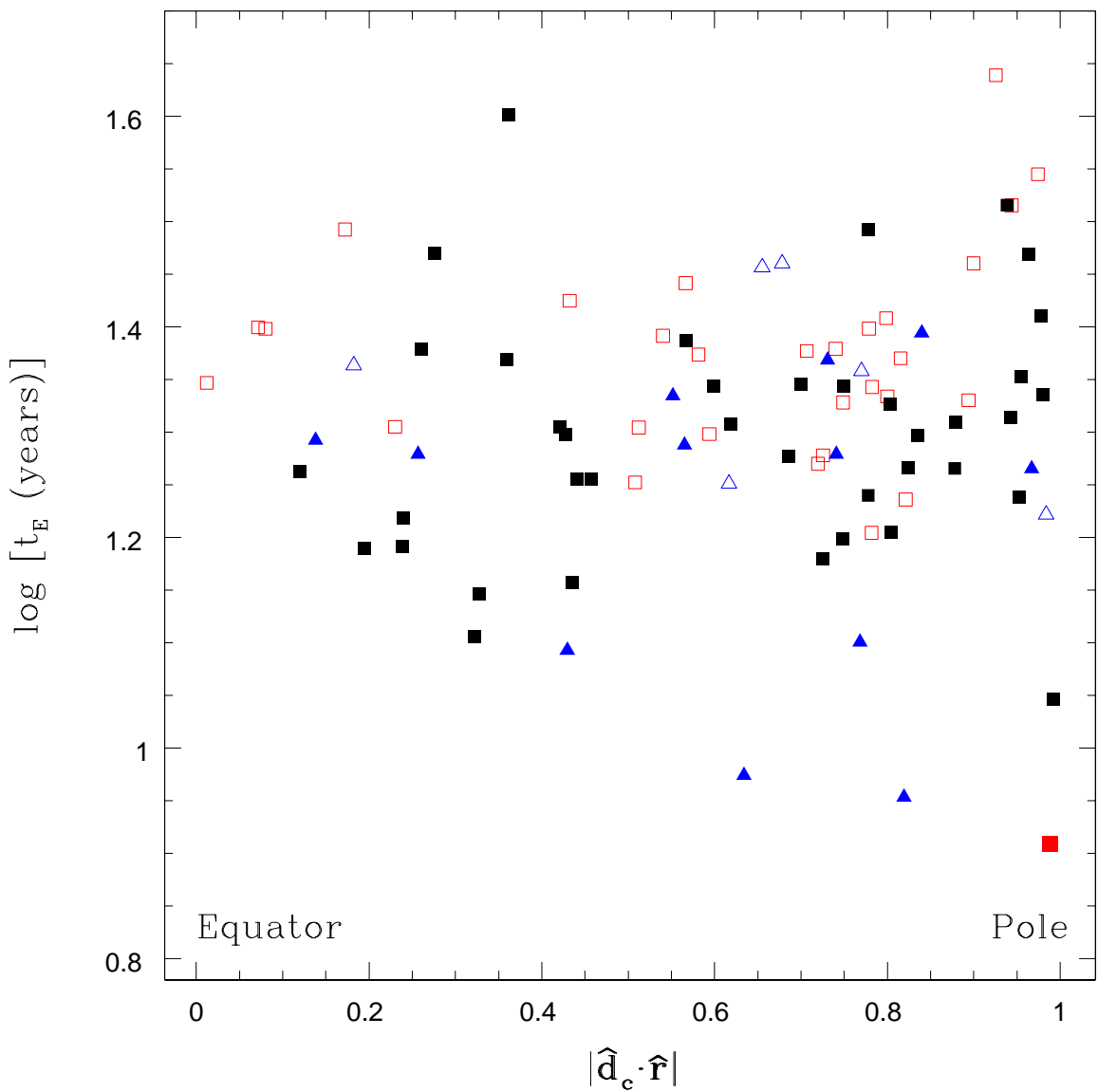}
\caption{\label{cmb} Einstein crossing timescale $t_E$ as a 
function of the position of the lens with respect to the CMB dipole location 
($\mathbf{\hat {d}_c \cdot \hat {r}}$). Filled symbols correspond to systems 
with spectroscopic redshifts for $z_l$, while open symbols are those with 
only estimates. Squares are used for typical  systems while triangles correspond 
to systems associated with groups/clusters with higher than usual internal 
velocities. The larger filled red square corresponds to Q~2237+0305. There is a weak 
trend with the location of the lens relative to the dipole.}
\end{center}
\end{figure}

Finally, we note that the dependence on the  projection of the CMB dipole velocity \citep{cmb09} means that microlensing 
variability should be enhanced along the equator of the dipole and suppressed along the poles. Figure \ref{cmb} 
shows the distribution of the Einstein crossing timescales 
as a function of the cosine between the dipole and the lens positions $\mathbf{\hat{d}_c \cdot \hat{r}}$. 
While there is significant scatter due to the differences in redshift and sizes, a modest trend is present for the systems with 
spectroscopic redshifts (filled squares), and excluding systems lensed by galaxy groups or 
cluster (triangles) since their higher velocity dispersions overwhelm the dipole contribution. Since 
the CMB dipole contribution is only one of three similar velocity scales in Equation \ref{vel_ef} it is never a dominant 
effect. At present, the CMB dipole is useful as a prior on microlensing models and to help select lenses with 
shorter variability timescales, but, in theory, microlensing could be used to confirm the kinematic origin for the dipole.

\bigskip

\noindent Acknowledgments:
We would like to thank M. Dietrich and J. Blackburne for helpful comments on 
the manuscript, and we thank J. Tinker for kindly providing the galaxy velocity 
dispersions. A.M.M. acknowledges the support of Generalitat Valenciana, 
grant APOSTD/2010/030. C.S.K. is supported by NSF grant AST-0708082 and AST-1004756.


\clearpage

\clearpage

\clearpage

\clearpage

\clearpage

\clearpage

\clearpage

\clearpage

\clearpage

\clearpage


\input {Tab1.tex}

\input {Tab2.tex}

\end{document}

%% file: Tab1.tex
\begin{deluxetable}{lccccccrrcccc}
\tabletypesize{\scriptsize}
\rotate
\tablecaption{\small \label {table1} Length and timescales for the selected sample of lensed quasars.}
\tablewidth{0pt}
\tablehead{
\colhead{Lens} & \colhead{$z_s$\tablenotemark{~}} & \colhead{$z_l$} &\colhead{$m_s$} & \multicolumn{1}{c}{$R_E$} & \multicolumn{1}{c}{$R_S$} & \multicolumn{1}{c}{$R_{\mathrm{BLR}}$} & \multicolumn{1}{c}{$t_E$} 
&  \multicolumn{1}{c} {$t_S$} &  \colhead{$R_S/R_E$}& \multicolumn{1}{c}{$M_{BH}$} & \colhead{Line} & \colhead{Ref}\\
& & & & ($10^{16}$ cm) &($10^{15}$ cm)&($10^{17}$ cm)&(years) &(years) & & ($10^9 M_{\odot}$) & & }
\startdata
 HE0047$-$1756     &     1.66	&       0.41    &       16.52   &      3.12    &      1.25   &       	1.46	   &      18.45    &      0.74    &  0.040  & 1.38 &   (CIV,a)               &  	    59, 90     \\      
PMNJ0134$-$0931	  &	2.22	&	0.77	&	18.96	&      2.29    &      0.09   &      	0.12	   &      31.05    &      0.13    &  0.004  &	-    &     -        	     & 	    19, 34, 59, 84     \\     
Q0142$-$100	  &	2.72	&	0.49	&	16.47	&      2.84    &      2.67   &      	4.72	   &      22.06    &      2.07    &  0.094  &	3.89 &  (CIV, b)      	     & 	    13, 74, 76, 77     \\     
QJ0158$-$4325	  &	1.29	&	0.32	&	17.39	&      3.41    &      1.62   &      	1.71	   &      17.99    &      0.86    &  0.048  &	-    &     -         	     & 	    16                 \\     
B0218$+$357	  &	0.96	&	0.68	&	19.28	&      1.54    &      0.30   &      	0.28	   &      20.30    &      0.40    &  0.019  &	-    &     -         	     & 	    6, 81, 58          \\     
HE0230$-$2130$^{\dag}$&	2.16	&	0.53	&	18.00	&      2.78    &      0.55   &      	0.76	   &      21.60    &      0.43    &  0.020  &	-    &     -         	     & 	    59, 88             \\     
SDSS0246$-$0825	  &	1.69	&	0.72	&	16.97	&      2.24    &      1.17   &      	1.36	   &      24.36    &      1.27    &  0.052  &	0.17 &   (CIV, b)     	     & 	    13, 25             \\     
MG0414+0534	  &	2.64	&	0.96	&	19.62	&      2.04    &      0.19   &      	0.30	   &      23.94    &      0.22    &  0.009  &	1.82 &  (H$\beta$, a) 	     & 	    79                 \\     
HE0435$-$1223	  &	1.69	&	0.46	&	16.84	&      2.94    &      0.76   &      	0.88	   &      18.30    &      0.47    &  0.026  &	0.50 &   (CIV, a)     	     & 	    59, 54, 89         \\     
HE0512$-$3329	  &	1.57	&       (0.93)	&	16.27	&      1.76    &      1.81   &      	2.07	   &      25.10    &      2.58    &  0.103  &	-    &     -          	     & 	    17                 \\     
B0712$+$472$^{\dag}$&	1.34	&	0.41	&	22.42	&      2.97    &      0.05   &      	0.05	   &      19.02    &      0.03    &  0.002  &	0.07 &  (MgII, a)     	     & 	    15                 \\     
SDSSJ0746$+$4403  &	2.00	&	0.51	&	18.34	&      2.82    &      0.63   &      	0.74	   &      23.39    &      0.52    &  0.022  &	-    &     -          	     & 	    28, 36             \\     
SDSSJ0806$+$2006  &	1.54	&	0.54	&	18.88	&      2.64    &      0.84   &      	0.93	   &      22.06    &      0.70    &  0.032  &	0.   &     -          	     & 	    13, 26             \\     
HS0810$+$2554	  &	1.50	&      ``0.89"	&	15.03	&      1.97    &      0.98   &      	1.07	   &      23.64    &      1.17    &  0.050  &	0.11 &  (CIV, a)     	     & 	    69                 \\     
HS0818$+$1227	  &	3.11	&	0.39	&	18.54	&      3.07    &      0.69   &      	1.27	   &      18.91    &      0.42    &  0.023  &	-    &     -         	     & 	    18                 \\     
SDSSJ0819$+$5356  &	2.24	&	0.29	&	18.51	&      3.70    &      0.53   &      	0.75	   &      14.00    &      0.20    &  0.014  &	-    &     -         	     & 	    29                 \\     
SDSSJ0820$+$0812  &	2.02	&	0.80	&	19.05	&      2.20    &      0.88   &      	1.08	   &      22.16    &      0.89    &  0.040  &	-    &     -         	     & 	    32                 \\     
APM08279$+$5255	  &	3.87	&	1.06	&	14.55	&      1.88    &      0.99   &      	2.15	   &      39.91    &      2.11    &  0.053  &	-    &     -         	     & 	    30, 39             \\     
SDSSJ0832$+$0404  &	1.12	&	0.66	&	18.89	&      1.90    &      0.99   &      	0.98	   &      15.82    &      0.83    &  0.052  &	-    &     -         	     & 	    64                 \\     
B0850$+$054	  &  1.14/3.93  &	0.59	&	23.15	&      2.14    &      0.12   &      	0.11	   &      25.61    &      0.15    &  0.006  &	-    &     -         	     & 	    3, 47              \\     
SDSSJ0903$+$5028  &	3.61	&	0.39	&	18.51	&      2.95    &      0.95   &      	1.95	   &      18.00    &      0.58    &  0.032  &	-    &     -         	     & 	    33                 \\     
SDSSJ0904$+$1512  &	1.83	&  [0.19]\{0.54\}&	17.51	&      2.72    &      1.29   &      	1.50	   &      25.03    &      1.19    &  0.047  &	-    &     -         	     & 	    36                 \\     
RXJ0911$+$0551$^{\dag}$	&2.80	&	0.77	&	17.39	&      2.29    &      0.59   &      	1.00	   &      24.78    &      0.64    &  0.026  &	0.80 &  (CIV, a)      	     & 	    5, 37              \\     
SBS0909$+$523	  &	1.38	&	0.83	&	15.65	&      1.79    &      4.46   &      	5.01	   &      20.17    &      5.02    &  0.249  &	1.95 &  (H$\beta$,b)  	     & 	    45                 \\     
SDSSJ0924$+$0219  &	1.52	&	0.39	&	18.18	&      3.16    &      0.61   &      	0.66	   &      20.39    &      0.39    &  0.019  &	0.11 &  (MgII, a)     	     & 	    11, 23, 59         \\     
SDSSJ0946$+$1835  &	4.80	&	0.38	&	18.74	&      2.65    &      0.82   &      	2.09	   &      19.80    &      0.61    &  0.031  &	-    &      -         	     &      48                 \\     
FBQ0951$+$2635	  &	1.24	&	0.26	&	16.39	&      3.92    &      3.93   &      	4.23	   &      17.37    &      1.74    &  0.100  &	0.89 &  (MgII, a)     	     & 	    13, 71             \\     
BRI0952$-$0115	  &	4.50	&	0.63	&	18.27	&      2.23    &      0.65   &      	1.52	   &      32.77    &      0.95    &  0.029  &	1.39 &  (CIV, a)      	     & 	    13, 49, 75         \\     
Q0957$+$561$^{\dag}$&	1.41	&	0.36	&	15.99	&      3.25    &      3.12   &      	3.48	   &      12.39    &      1.19    &  0.096  &	2.01 &  (CIV, a)      	     & 	    80, 91             \\     
SDSSJ1001$+$5027$^{\dag}$&1.84	&      ``0.87"	&	17.31	&      3.08    &      1.61   &      	1.86	   &      17.88    &      0.93    &  0.052  &	-     &    -          	      &	    63                 \\     
J1004$+$1229	  &	2.65	&	0.95*	&	19.65	&      2.06    &      0.70   &      	1.15	   &      28.86    &      0.98    &  0.034  &	-     &    -         	      &	    41                 \\     
SDSSJ1004$+$4112$^{\dag}$&1.73	&	0.68	&	17.53	&      2.35    &      0.69   &      	0.80	   &       9.42    &      0.28    &  0.029  &	2.02  &  (CIV, a)     	      &	    24, 61             \\     
SDSSJ1011$+$0143  &	2.70	&	0.33	&	22.43	&      3.41    &      0.06   &      	0.10	   &      17.32    &      0.03    &  0.002  &	-     &    -         	      &	    4                  \\     
SDSSJ1021$+$4913  &	1.72	&     ``0.94"	&	18.97	&      2.06    &      0.74   &      	0.86	   &      24.63    &      0.89    &  0.036  &	-     &    -         	      &	    68                 \\     
LBQS1009$-$0252	  &	2.74	&	0.87	&	17.80	&      2.16    &      1.49   &      	2.59	   &      29.45    &      2.04    &  0.069  &	1.64  &  (CIV, a)    	      &	    21, 59             \\     
Q1017$-$207	  &	2.55	&      (0.78)	&	16.78	&      2.28    &      2.00   &      	3.33	   &      32.76    &      2.87    &  0.088  &	1.68  &  (CIV, a)    	      &	    9, 38              \\     
SDSSJ1029$+$2623$^{\dag}$&2.20	&	0.55	&	18.55	&      2.73    &      0.45   &      	0.62	   &       8.98    &      0.15    &  0.016  &	-     &    -         	      &	    27                 \\     
B1030$+$074	  &	1.54	&	0.60	&	19.41	&      2.46    &      0.83   &      	0.91	   &      22.51    &      0.76    &  0.034  &	0.35  &  (MgII, a)   	      &	    15                 \\     
SDSSJ1054$+$2733  &	1.45	&      [0.23]	&	16.81	&      4.13    &      1.31   &      	1.44	   &      17.22    &      0.55    &  0.032  &	-     &     -         	      &	    36                 \\     
SDSSJ1055$+$4628  &	1.25	&  [0.39]\{0.38\}&	18.76	&      3.04    &      1.02   &      	1.05	   &      19.87    &      0.67    &  0.033  &	-     &     -        	      &	    36                 \\     
HE1104$-$1805	  &	2.32	&	0.73	&	16.17	&      2.36    &      2.43   &      	3.75	   &      21.66    &      2.23    &  0.103  &	2.37  & (CIV, a)     	      &	    44, 86             \\     
PG1115$+$080$^{\dag}$&	1.72	&	 0.31	&	15.62	&      3.62    &      1.19   &      	1.41	   &      18.43    &      0.61    &  0.033  &	0.92  & (CIV, a)      	      &	    20, 8, 78          \\     
RXJ1131$-$1231 &	0.66	&	0.29	&	16.74	&      2.50    &      0.64   &      	0.45	   &      11.13    &      0.28    &  0.026  &	0.06  & (H$\beta$, a) 	      &	    73                 \\     
SDSSJ1131$+$1915  &	2.92	&      [0.32]	&	18.00	&      3.41    &      0.78   &      	1.38	   &      21.40    &      0.49    &  0.023  &	-     &    -          	      &	    36                 \\     
SDSSJ1138$+$0314 &	2.44	&	0.45	&	18.43	&      3.00    &      0.44   &      	0.67	   &      25.74    &      0.38    &  0.015  &	0.05  & (CIV, b)      	      &	    12                 \\     
SDSSJ1155$+$6346 &	2.89	&	0.18	&	17.67	&      4.49    &      1.88   &      	3.42	   &      12.75    &      0.53    &  0.042  &	-     &    -                  &	    67                 \\     
B1152$+$200	  &	1.02	&	0.44	&	16.53	&      2.52    &      2.72   &      	2.74	   &      18.43    &      1.99    &  0.108  &	-     &    -          	      &	    57                 \\     
SDSSJ1206$+$4332$^{\dag}$&1.79	&      ``0.85"	&	18.47	&      3.11    &      0.71   &      	0.82	   &      17.83    &      0.40    &  0.023  &	-     &    -          	      &	    63                 \\     
Q1208$+$101	  &	3.80	& ``1.33", 1.14*&	16.96	&      1.27    &      1.84   &      	4.04	   &      43.54    &      6.33    &  0.145  &	-     &    -          	      &	    2, 72              \\     
SDSSJ1216$+$3529  &	2.01	&	[0.55]	&	19.08	&      2.72    &      0.56   &      	0.67	   &      23.83    &      0.49    &  0.021  &	-     &    -          	      &	    64                 \\     
SDSSJ1226$-$0006  &	1.12	&	0.52	&	18.30	&      2.35    &      0.86   &      	0.85	   &      20.60    &      0.76    &  0.037  &	0.68  & (MgII, a)      	      &	    12                 \\     
SDSSJ1251$+$2935  &	0.80	&	0.41	&	18.85	&      2.24    &      0.30   &      	0.26	   &      15.12    &      0.20    &  0.013  &	-     &    -          	      &	    35                 \\     
SDSSJ1254$+$2235  &	3.63	&      \{0.2\}	&	19.19	&      3.96    &      0.46   &      	0.93	   &      16.01    &      0.19    &  0.012  &	-     &    -          	      &	    29                 \\     
SDSSJ1258$+$1657  &	2.70	&	[0.4]	&	18.74	&      3.12    &      1.04   &      	1.78	   &      23.45    &      0.78    &  0.033  &	-     &    -          	      &	    29                 \\     
SDSSJ1304$+$2001  &	2.18	&  [0.32]\{0.46\}&	18.45	&      2.99    &      0.77   &      	1.07	   &      22.03    &      0.57    &  0.026  &	-     &    -         	      &	    36                 \\     
SDSSJ1313$+$5151  &	1.88	&	0.19	&	17.72	&      4.56    &      1.80   &      	2.05	   &      14.37    &      0.57    &  0.039  &	-     &    -         	      &	    60                 \\     
SDSSJ1322$+$1052  &	1.72	&  ``0.88"[0.55]&	18.24	&      2.64    &      0.90   &      	1.05	   &      21.58    &      0.74    &  0.034  &	-     &    -           	      &	    64                 \\     
SDSSJ1330$+$1810$^{\dag}$ &1.40	&	0.37	&	18.34	&      3.17    &      0.49   &      	0.50	   &      19.02    &      0.29    &  0.015  &	-     &    -         	      &	    65                 \\     
SDSSJ1332$+$0347  &	1.45	&	0.19	&	18.70	&      4.53    &      0.97   &      	1.05	   &      16.03    &      0.34    &  0.021  &	-     &    -         	      &	    55                 \\     
LBQS1333$+$0113	  &	1.57	&	0.44	&	17.26	&      2.97    &      1.84   &      	2.10	   &      21.19    &      1.31    &  0.062  &	1.55  &  (MgII, a)    	      &	    12, 62             \\     
SDSSJ1339$+$1310  &	2.24	&	[0.4]	&	18.71	&      3.19    &      0.80   &      	1.16	   &      21.29    &      0.54    &  0.025  &	-     &    -        	      &	    29                 \\     
SDSSJ1349$+$1227  &	1.72	& [0.63]\{0.66\}&	17.44	&      2.39    &      2.48   &      	3.02	   &      18.97    &      1.97    &  0.104  &	-     &    -        	      &	    36                 \\     
SDSSJ1353$+$1138  &	1.63	&	0.3*	&	16.48	&      3.66    &      2.80   &      	3.33	   &      18.62    &      1.43    &  0.077  &	-     &    -        	      &	    26                 \\     
Q1355$-$2257	  &	1.37	&	0.48*	&	16.94	&      2.92    &      2.18   &      	2.67	   &      23.94    &      1.79    &  0.075  &	-     &    -        	      &	    51, 59             \\     
B1359+154$^{\dag}$ &    3.24    &      ``0.9''  &       22.62   &      2.09    &      0.05   &       	0.09	   &      28.85    &      0.07    &  0.002  &   -     &    -        	      &	    57, 70             \\     
SDSSJ1400$+$3134  &	3.32	&      \{0.8\}	&	19.92	&      2.20    &      0.43   &      	0.80	   &      27.64    &      0.53    &  0.019  &	-     &    -        	      &	    29                 \\     
SDSSJ1406$+$6126  &	2.13	&	0.27	&	18.88	&      3.87    &      0.78   &      	1.06	   &      15.55    &      0.31    &  0.020  &	-     &    -        	      &	    28                 \\     
H1413$+$117$^{\dag}$&	2.55	&     ``0.94''	&	16.44	&      2.07    &      1.05   &      	1.71	   &      28.62    &      1.46    &  0.051  &	0.26  & (CIV, a)     	      &	    46                 \\     
B1422$+$231$^{\dag}$&	3.62	&	0.34	&	14.81	&      3.12    &      2.29   &      	4.88	   &      19.39    &      1.42    &  0.073  &	4.79  & (CIV, a)     	      &	    66, 78             \\     
SDSSJ1455$+$1447  &	1.42	& [0.27]\{0.53\}&	18.22	&      2.59    &      0.69   &      	0.73	   &      20.16    &      0.53    &  0.027  &	-     &    -        	      &	    36                 \\     
SBS1520$+$530$^{\dag}$&	1.86	&	0.72*	&	17.61	&      2.31    &      1.77   &      	2.02	   &      23.09    &      1.77    &  0.077  &	0.88  & (CIV, a)     	      &	    1, 7               \\     
SDSSJ1524$+$4409  &	1.21	&	0.32	&	18.76	&      3.33    &      0.69   &      	0.69	   &      16.54    &      0.34    &  0.021  &	-     &    -         	      &	    64                 \\     
B1600$+$434$^{\dag}$&	1.59	&	0.41	&	20.87	&      3.10    &      0.34   &      	0.37	   &      19.61    &      0.22    &  0.011  &	0.10   & (MgII, a)     	      &	    15                 \\     
SDSSJ1620$+$1203  &	1.16	&	0.40	&	19.10	&      2.87    &      0.95   &      	0.95	   &      15.47    &      0.51    &  0.033  &	-     &    -         	      &	    36                 \\     
PMNJ1632$-$0033	  &	3.42	&	1.16*	&	20.68	&      1.83    &      0.47   &      	0.91	   &      31.08    &      0.80    &  0.026  &	0.39  & (CIV, a)      	      &	    59, 85             \\     
FBQ1633$+$3134	  &	1.52	&	0.68*	&	16.59	&      2.24    &      2.82   &      	3.23	   &      25.01    &      3.14    &  0.126  &	1.76  & (CIV, a)      	      &	    50                 \\     
SDSSJ1650$+$4251  &	1.54	&	0.58*	&	16.98	&      2.51    &      1.81   &      	2.06	   &      22.23    &      1.60    &  0.072  &	-     &   -          	      &	    52                 \\     
PKS1830$-$211	  &	2.51	&	0.89	&	22.27	&      2.13    &      0.12   &      	0.17	   &      29.52    &      0.16    &  0.006  &	-     &   -          	      &	    43, 82             \\     
PMNJ1838$-$3427	  &	2.78	&      [0.36]	&	19.10	&      3.26    &      1.28   &      	2.23	   &      20.19    &      0.79    &  0.039  &	-     &   -          	      &	    83                 \\     
MG2016$+$112$^{\dag}$&	3.27	&	1.01	&	21.51	&      1.97    &      0.17   &      	0.30	   &      23.37    &      0.20    &  0.009  &	-     &   -          	      &	    40, 42             \\     
WFI2026$-$4536	  &	2.23	&     ``1.04"	&	16.18	&      2.13    &      1.12   &      	1.62	   &      26.60    &      1.40    &  0.053  &	-     &   -           	      &	    53                 \\     
WFI2033$-$4723	  &	1.66	&	0.66	&	17.59	&      2.37    &      0.71   &      	0.81	   &      19.86    &      0.60    &  0.030  &	-     &   -           	      &	    53, 59             \\     
B2045$+$265$^{\dag}$&	1.28	&	0.87	&	22.02	&      1.57    &      0.02   &      	0.02	   &      12.61    &      0.02    &  0.001  &	0.01  &  (MgII, a)            &	    14                 \\     
HE2149$-$2745$^{\dag}$&	2.03	&      0.60	&	16.29	&      2.86    &      3.08   &      	3.99	   &      22.81    &      2.45    &  0.108  &	6.62  &  (CIV, a)             &	    13, 87             \\     
Q2237$+$030	  &	1.69	&	0.04	&	15.16	&      9.90    &      2.84   &      	3.44	   &       8.11    &      0.23    &  0.027  &	0.47  &  (CIV, b)             &	    22                 \\     
PSS2322$+$1944	  & 	4.12	&      ``1.23"	&	17.92	&      1.81    &      0.88   &      	1.93	   &      35.08    &      1.70    &  0.049  &	2.36  &   lower limit, a      &	    10                 \\     
SDSSJ2343$-$0050$^{\dag}$ &0.79	&	0.30**	&	20.10	&      2.82    &      0.42   &  	0.36	   &      16.66    &      0.25    &  0.015  &	-     &     -                 &	    31                 \\ 
\enddata
\tablenotetext{~}{\\ 
\vspace{0.05cm}
(1) \cite{matt08}; 
(2) \cite{bahcall92}; 
(3) \cite{biggs03}; 
(4) \cite{bolton06}; 
(5) \cite{burud98}; 
(6) \cite{carilli93}; 
(7) \cite{Chavushyan}; 
(8) \cite{Christian87}; 
(9) \cite{Claeskens96}; 
(10) \cite{Cox02}; 
(11) \cite{Eigenbrod06a}; 
(12) \cite{Eigenbrod06b}; 
(13) \cite{Eigenbrod07}; 
(14) \cite{Fassnacht99}; 
(15) \cite{Fassnacht98}; 
(16) \cite{Faure09}; 
(17) \cite{Gregg2000};
(18) \cite{Hagen00};
(19) \cite{Hall02}; 
(20) \cite{HenryHeasley86};
(21) \cite{hewett94};	
(22) \cite{Huchra85}; 
(23) \cite{Inada03a}; 
(24) \cite{Inada03b}; 
(25) \cite{Inada05}; 
(26) \cite{Inada06a}; 
(27) \cite{Inada06b}; 
(28) \cite{Inada07}; 
(29) \cite{Inada09}; 
(30) \cite{Irwin98}; 
(31) \cite{Jackson08}; 
(32) \cite{Jackson09}; 
(33) \cite{Johnston03};	
(34) \cite{Kanekar03}; 
(35) \cite{kayo07}; 
(36) \cite{kayo10}; 
(37) \cite{kneib00};
(38) \cite {CK00};
(39) \cite{kondo06}; 
(40) \cite{leon02}; 
(41) \cite{Lacy02}; 
(42) \cite{Lawrence84}; 
(43) \cite{Lidman99}; 
(44) \cite{Lidman00}; 
(45) \cite{Lubin00}; 
(46) \cite{Magain88}; 
(47) \cite{McKean04}; 
(48) \cite{McGreer10}; 
(49) \cite{mm_irwin92}; 
(50) \cite{morgan01}; 
(51) \cite{Morgan03a}; 
(52) \cite{Morgan03b};
(53) \cite{Morgan04};
(54) \cite{morgan2005};
(55) \cite{Morokuma07}; 
(56) \cite{pp2001}; 
(57) \cite{mayers99}; 
(58) \cite{ODea92}; 
(59) \cite{Ofek06}; 
(60) \cite{Ofek07}; 
(61) \cite{Oguri04a}; 
(62) \cite{Oguri04b}; 
(63) \cite{Oguri05}; 
(64) \cite{Oguri08a}; 
(65) \cite{Oguri08b}; 
(66) \cite{Patnaik92}; 
(67) \cite{Pindor04}; 
(68) \cite{Pindor06}; 
(69) \cite{Reimers02}; 
(70) \cite{Rusin00}; 
(71) \cite{Schechter98}; 
(72) \cite{Siemiginowska98}; 
(73) \cite{Sluse03}; 
(74) \cite{Smette92}; 
(75) \cite{SL96}; 
(76) \cite{Surdej87}; 
(77) \cite{Surdej88}; 
(78) \cite{Tonry98}; 
(79) \cite{tonry_kochanek99}; 
(80) \cite{Walsh79}; 
(81) \cite{Wiklind95}; 
(82) \cite{Wiklind96};
(83) \cite{Winn00}; 
(84) \cite{Winn02a}; 
(85) \cite{Winnb}; 
(86) \cite{Lutz93};	
(87) \cite{Wisotzki98}; 
(88) \cite{Wisotzki99}; 
(89) \cite{Wisotzki02}; 
(90) \cite{Wisotzki04}; 
(91) \cite{Young80};\\
Notes:\\
$^{\dag}$ Lensing by a group or cluster of galaxies\\
a-\cite{peng06} \\
b-\cite{Roberto10}\\
(~) based on the FP method \\
 $[~]$ based on the FJ relation\\
\{ \} based on color measurments\\
$^\ast$ $^\ast$ based on spectrum features\\
$^\ast$ based on the absorption line spectrum, but not completely identified \\
`` " estimated from the image separation and $z_l$ probability distriburion \cite{ofek03}}
\end{deluxetable}

%% file: Tab2.tex
\def\foo{\hphantom{$-$}}
\def\fff{\hphantom{1}}
\begin{deluxetable}{ccc}
\tabletypesize{\small}
\tablecaption{\small \label {coef} Fitted coefficients to 
the $\sigma_{\mathrm{pec}}-z$ relationship (Equation \ref{spec}). 
}
\tablewidth{0pt}
\setlength{\tabcolsep}{13.2pt}
\tablehead{
\colhead{$z$-bin} & \colhead{$a$} & \colhead{$b$}}
\startdata
$[0.0502, 0.2162]$ &  \foo0.32   &  2.43 \\
$[0.2162, 0.4085]$ &  $-0.31$    &  2.48 \\
$[0.4085, 0.7132]$ &  $-0.05$    &  2.45 \\
$[0.7132, 0.9841]$ &  $-0.13$    &  2.47 \\
$[0.9841, 1.5342]$ &  $-0.57$    &  2.60 \\
\enddata
\tablenotetext{~}{$\log_{10}\left(\sigma_{\mathrm{pec}}/\mathrm{km \  s}^{-1}\right)= a~\log_{10}(1+z)+b$}
\end{deluxetable}